\documentclass[aps,pra,10pt, showpacs,twocolumn,superscriptaddress]{revtex4-2} 

\usepackage[utf8]{inputenc}
\usepackage{physics}
\usepackage{graphicx}
\usepackage{epstopdf}
\usepackage{setspace}
\usepackage{bbold}
\usepackage{comment}
\usepackage{color}
\usepackage{soul}
\usepackage[abs]{overpic}
\usepackage{amsmath}
\usepackage{lipsum}
\usepackage{placeins}
\usepackage{svg}
\usepackage[abs]{overpic}
\usepackage{float}
\usepackage[export]{adjustbox}
\usepackage{amsmath}
\usepackage{nicefrac}
\usepackage{cleveref}
\usepackage{layouts}
\usepackage{subcaption}

\usepackage[font=small, labelsep=period,
   justification=Justified,
   format=plain]{caption} % 'format=plain' avoids hanging indentation
\usepackage{ragged2e}
\graphicspath{ {./images/} }
\begin{document}

\title{Frequency Attraction and Repulsion in Lasers with Complex Mutual Coupling}
\author{Amit Pando}
\altaffiliation[These authors contributed equally to this work]{}

\author{Tomer Hacohen}
\altaffiliation[These authors contributed equally to this work]{}

\author{Eran Bernstein}
\author{Victor Shelukin}
\author{Asher Friesem}
\author{Nir Davidson}
\affiliation{Department of Physics of Complex Systems, Weizmann Institute of Science, Rehovot 7610001, Israel}
\begin{abstract}
We experimentally investigate laser synchronization under complex mutual coupling, featuring both dissipative and dispersive components. Using a reconfigurable digital degenerate cavity laser with precise control over system parameters, we show that complex coupling can induce both frequency attraction and repulsion, in contrast to common models of coupled oscillators. Furthermore, we show that these frequency shifts can be exploited to significantly enhance global synchronization in multi-laser arrays. Comparison with numerical and theoretical results indicate that amplitude dynamics of complex coupled lasers play an important role in the observed phenomena. Our results establish a versatile platform for exploring non-Hermitian physics and topological states in tailored laser networks.
\end{abstract}
\maketitle

\section{Introduction}
The study of synchronization of coupled oscillators has been ongoing for many centuries using clocks\cite{huygens1895letters}, Josephson junction arrays\cite{trees2005synchronization,wiesenfeld1998frequency}, coupled laser arrays\cite{pando2024synchronization,nixon2011synchronized}, swarms of fireflies\cite{ramirez2017fireflies} and human crowds\cite{shahal2020synchronization}. 
Generally, synchronization is the transition of interacting elements into a stable collective state. Accordingly, it is commonly understood that synchronization involves dissipation, which is required to reduce the available phase space of the system. In physical coupled mode systems such as laser arrays, Josephson junctions, and polariton condensates, synchronization can be viewed as the transfer of energy from incoherent states into a single coherent mode. In this context, dissipation is essential because purely dispersive interactions cannot remove energy from undesired modes.

Complex coupling, with both dissipative and dispersive components is involved in coupled biological oscillators\cite{sakaguchi1988mutual}, semiconductor lasers\cite{winful1988stability,kozyreff2001dynamics} and driven-dissipative condensates\cite{moroney2021synchronization}. Several theoretical and experimental works in recent years have shown that coupled laser arrays with complex coupling can have nontrivial topology and support topological states, leading to improved coherence\cite{harari2018topological,dikopoltsev2021topological,bandres2018topological,moroney2023synchronization,amelio2020theory,zapletal2020long}. In the temporal domain, complex coupling between consecutive mode-locked pulses has similarly been shown to induce spontaneous frequency shifts and drive the formation of multistable Kuramoto splay states \cite{seidel_multistable_2025}. Nevertheless, it is unclear if and how complex coupling affects synchronization.

In this work, we experimentally develop and examine a system of two coupled lasers with precisely controlled detuning and coupling in order to qualitatively study the effects of complex coupling on synchronization (phase locking). By measuring the beating frequency of the two lasers we show that complex coupling induces a mixture of dissipative and dispersive dynamics that can result in either frequency attraction or repulsion, depending on the coupling rate and phase. We also demonstrate that the frequency of the synchronized state can be tuned by changing the coupling phase. In all of these, amplitude dynamics induced by complex coupling play an important role. Then, we show that the synchronization of a system consisting of three lasers can be significantly improved with complex coupling. Our methods could be used in the future to explore the effects of dispersion and topology on coherence in large coupled laser systems\cite{manannikov2026klein}, and improve synchronization in tailored laser networks\cite{vigne2026tailoring}.

\begin{figure}[h]
    \centering
    \includegraphics[trim=5.8cm 0cm 5.7cm 0cm,clip,width=1\linewidth]{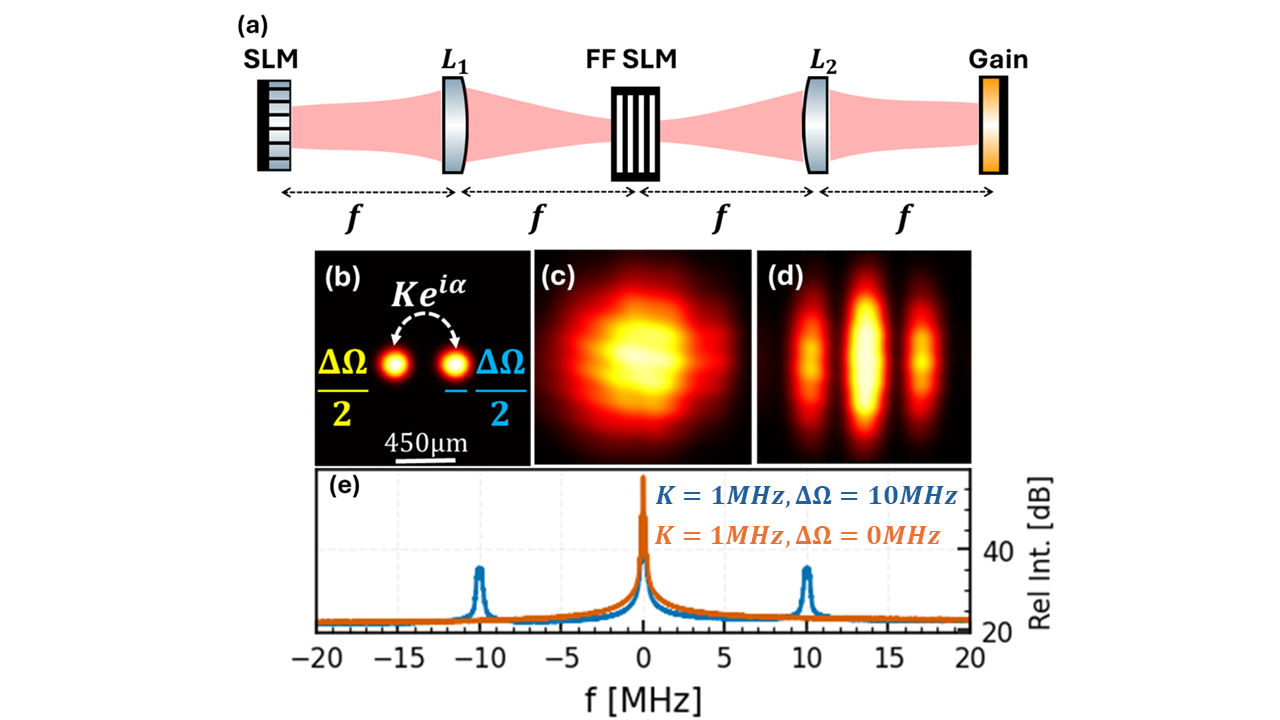}
    \caption{\textbf{(a)}: Experimental arrangement of a degenerate cavity laser with two intra-cavity lenses and two intra-cavity spatial light modulators (SLMs) in the near field (NF) and far field (FF) planes of the cavity. \textbf{(b):}The NF SLM defines two Gaussian laser spots, detuned by $\pm\frac{1}{2}\Delta\Omega$. \textbf{(c):}When the lasers are uncoupled they are incoherent, a broad Gaussian intensity distribution is detected in their common far field. \textbf{(d):} When the lasers are coupled, interference fringes are detected in the far field intensity distribution. \textbf{(e):} The resulting beating spectrum of the two lasers detected by a photodiode, with detuning (blue) and without (orange).}
    \label{fig:exp_setup}
\end{figure}

\section{Experimental Arrangement}
Our experimental arrangement, schematically shown in Fig. \ref{fig:exp_setup}(a) and described in detail in \cite{Supplemental}, consists of a digital degenerate cavity laser (DDCL) \cite{arnaud1969degenerate,cao2019complex,tradonsky2021high}. It includes an intra-cavity $4f$ telescope, a $3\text{mm}$ thick Nd:YVO4 gain medium lasing at $\lambda=1.06\mu\text{m}$, and two spatial light modulators (SLM) with pixel pitch of $8\mu\text{m}$, placed in the near field (NF) and far field (FF) planes of the cavity. The laser is pumped by an external diode laser ($\lambda_{pump}=808 n\text{m}$) with quasi-CW pulses of $400\mu\text{s}$ with a $4$Hz repetition rate to minimize local distortions of the cavity caused by thermal effects \cite{gadasi2023improved}. We manipulate the phase of adjacent blocks of $2\times2$ pixels to generate effective phase and amplitude masks \cite{tradonsky2021high}. An amplitude mask is displayed on the NF SLM (SLM in Fig. \ref{fig:exp_setup}(a)) to obtain Gaussian laser spots with waists of $140\mu\text{m}$ and distance between them $450\mu\text{m}$ (Fig. \ref{fig:exp_setup}(b)). The laser spots are measured in the $NF$ and $FF$ to evaluate $M^{2}\approx 1.1$. The NF SLM is also used to precisely control the natural frequency of the two lasers: by controlling the phase retardation of each pixel, so as to change the local cavity length with a resolution of $\frac{\lambda}{256}$ and hence change the resonant frequency of each laser with a resolution of $\var\Omega = \frac{\nu_{FSR}}{256}$, where $\nu_{FSR} = \frac{c}{2l}\approx 76.6$MHz is the free spectral range of the cavity.
A beam sampler (not shown) is used to extract a small portion of the cavity light in order to detect the NF and FF intensity distributions, as well as measure the beating frequency of the two lasers.

The gain medium's bandwidth is much larger than $\nu_{FSR}$, resulting in many longitudinal modes lasing in the cavity. We measured the lasing bandwidth to be $\sim50$GHz (depending on exact lasing conditions), which are equivalent to hundreds of longitudinal modes\cite{chriki2018spatiotemporal}. The longitudinal modes in the system do not interact, and are incoherently averaged upon detection \cite{nixon2013observing,pal2020rapid}.

Coupling between the lasers is controlled (rate $K$ and phase $\alpha$) by means of a phase and amplitude grating displayed on the FF SLM. The exact properties of the grating can be easily calculated using simple Fourier calculations, shown in detail in Appendix A. When the lasers are uncoupled and are incoherent with each other, a broad Gaussian intensity distribution is observed in the FF (Fig. \ref{fig:exp_setup}(c)), whereas when they are coupled and coherent with each other, high contrast interference fringes appear (Fig. \ref{fig:exp_setup}(d)).

The beating frequency of the two lasers is detected with an amplified photodiode (Fig. \ref{fig:exp_setup}(e)) as detailed in Sec. II of \cite{Supplemental}. The obtained spectra are averaged over 30 shots. In all results shown below, the errorbars were negligible compared to the separation of datapoints and therefore omitted. Intra-cavity adaptive optics protocol was used on both SLMs to reduce extraneous coupling and detuning between the lasers which might result from slight misalignment and optical aberrations\cite{pando2023improved}.

\section{Effects of complex coupling}

\begin{figure}[h]
    \centering
    \includegraphics[width=1\linewidth]{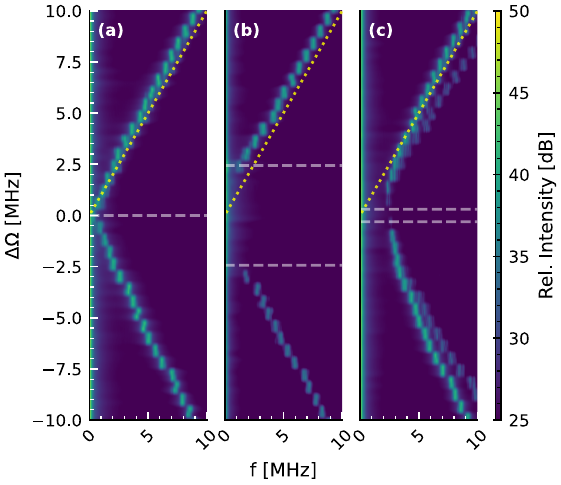}
    \caption{Measured beating spectrum as function of detuning $\Delta\Omega$ for \textbf{(a)} no coupling ($K=0$), \textbf{(b)} dissipative coupling ($K=1.2$MHz, $\alpha=0$) and \textbf{(c)} complex coupling ($K=1$MHz, $\alpha=0.45\pi$). Dashed lines represent the expected locking regions using the Sakaguchi-Kuramoto model, $\abs{\Delta\Omega_c} = 2K\cos\alpha$\cite{sakaguchi1988mutual}. The yellow dotted line illustrates a linear trend $\Delta\omega = \Delta\Omega$.}
    \label{fig_spec}
\end{figure}

To investigate the effects of complex coupling on synchronization (phase locking), we first consider a system of two lasers which are symmetrically detuned, $\Omega_{1,2} = \pm \frac{1}{2}\Delta\Omega$. Figure \ref{fig_spec} shows the measured beating spectrum of the two lasers as a function of relative detuning $\Delta\Omega$ for different coupling rates and phases. When the lasers are uncoupled ($K=0$, Fig. \ref{fig_spec}(a)) the measured beating frequency $\Delta\omega$ is equal to the applied detuning $\Delta\Omega$. When the two lasers are coupled with dissipative coupling ($K=1.2$MHz, $\alpha=0$, Fig. \ref{fig_spec}(b)) and complex coupling ($K=1$MHz, $\alpha=0.45\pi$, Fig. \ref{fig_spec}(c)), the lasers can phase lock as long as their detuning is smaller than some critical value $\Delta\Omega_c$. We observe that for $\Delta\Omega\gtrsim \Delta\Omega_c$, the measured frequency difference between the lasers $\Delta\omega$ diverges from the linear trend (shown in a yellow dotted line). In the case of dissipative coupling, the lasers undergo frequency attraction, such that $\Delta\omega < \Delta\Omega$, whereas in the case of complex coupling the two lasers undergo frequency repulsion such that $\Delta\omega > \Delta \Omega$. This repulsion is reminiscent of an avoided crossing observed in linear dispersive systems. In some measurements, a parasitic higher order transverse mode lases in the cavity, resulting in the additional beating signal visible in Fig. \ref{fig_spec}(c). Due to the degeneracy of the cavity, the additional mode does not interact with the fundamental mode in a meaningful way, and therefore we neglect it throughout this work.

We analyze the phase locking dynamics of the two lasers using the coupled laser rate equations (LRE) of class B lasers\cite{PhysRevLett.92.093905},
\begin{subequations}
\begin{align}
 &\dot{A}_i = \frac{1}{\tau_c}(G_i-\gamma_i)A_i + K_{ij}A_j\cos(\phi_j-\phi_i+\alpha_{ij}), \label{eqn_LRE_A}\\
 & \dot{\phi}_i = \Omega_i + K_{ij}\frac{A_j}{A_i}\sin(\phi_j-\phi_i+\alpha_{ij}),\label{eqn_LRE_phi}\\
 &\dot{G}_i = \frac{1}{\tau_f}[P_i-G_i(1+A_i^2)], \label{eqn_LRE_G}
\end{align}
\end{subequations}
where $A_i,\phi_i,G_i,\gamma_i,\Omega_i,P_i$ are amplitude, phase, gain, loss, detuning and pump rate the $i$-th laser, $\tau_c$ the cavity roundtrip time, $\tau_f$ the gain fluorescence lifetime, $K_{ij}$ the coupling rate between lasers $i$ and $j$ and $\alpha_{ij}$ is the coupling phase.  Throughout this paper we assume $P_i = P, \gamma_i =\gamma$ for all lasers, and consider only reciprocal coupling, $K_{ij} = K_{ji}, \alpha_{ij}=\alpha_{ji}
$.
The frequency difference of two mutually coupled lasers, $\Delta\omega \equiv \dot{\phi}_1-\dot{\phi}_2$ is given by

\begin{equation}\label{eqn_freq_lock}
    \begin{split}
            &\Delta\omega  = \Delta\Omega - K\cos\alpha(\frac{A_1}{A_2} + \frac{A_2}{A_1})\sin(\phi_1-\phi_2)\\ &-K\sin\alpha(\frac{A_1}{A_2}-\frac{A_2}{A_1})\cos(\phi_1-\phi_2),
    \end{split}
\end{equation}
where $\Delta\Omega \equiv \Omega_1-\Omega_2$, and the lasers are considered to be locked if $\Delta\omega =0$. The first term in the RHS of Eq. (\ref{eqn_freq_lock}) is the relative natural frequency detuning between the lasers, and the second and third terms are the dissipative and dispersive parts of the coupling. Notably, when the laser amplitudes are equal ($A_1=A_2$), the dispersive term vanishes and Eqs. (\ref{eqn_LRE_phi}) reduces to the Sakaguchi-Kuramoto model \cite{acebron2005kuramoto,sakaguchi1988mutual,sakaguchi1986soluble},
\begin{equation}\label{eqn_kuramoto_sakaguchi}
    \dot{\phi}_i = \Omega_i + K\sin(\phi_j-\phi_i+\alpha).
\end{equation}
The dispersive term also vanishes when the coupling is purely dissipative ($\alpha=0$), and Eq. (\ref{eqn_freq_lock}) reduces to the Adler equation for injection locking \cite{adler2006study,siegman1986lasers}. A similar decomposition can be done for the amplitude Eq. (\ref{eqn_LRE_A}), where dissipative coupling drives the system to a stable fixed point with equal amplitudes (see Sec. II in \cite{Supplemental}). In contrast, dispersive coupling drives the lasers to have amplitudes which are instantaneously different but identical on average. These competing effects result in nonlinear dynamics which are generally not analytically solvable, but can be studied perturbatively in the weak coupling regime (see Sec. VII in \cite{Supplemental}).

The dashed horizontal lines in Fig. \ref{fig_spec} denote the theoretically predicted critical detuning in the Sakaguchi-Kuramoto model below which the lasers are expected to be phase locked, 
\begin{equation}\label{eqn_critical_detuning}
    \abs{\Delta\Omega_c} = 2K\cos\alpha.
\end{equation}
As evident, our results in Fig. \ref{fig_spec} are in good agreement with Eq. (\ref{eqn_critical_detuning}). 

\begin{figure}[h]
    \centering
    \includegraphics[width=1\linewidth]{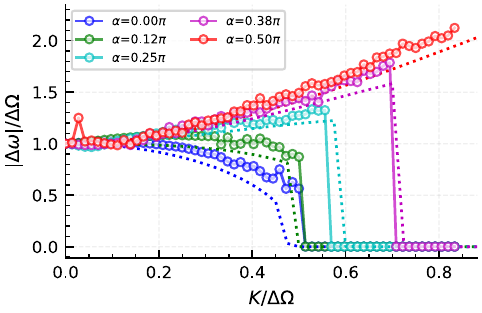}
    \caption{Measured normalized beating frequency $\abs{\Delta\omega}/\Delta\Omega$ as a function of normalized coupling $K/\Delta\Omega$ and coupling phase $\alpha$. $\Delta\Omega = 2.2$MHz for all measurements. It can be observed that complex coupling induces frequency repulsion.  Dotted curves denote numerical results of LRE simulations with matching parameters.}
    \label{fig_beat_alpha}
\end{figure}

 We further investigate the frequency attraction and repulsion effects of complex coupling by holding $\Delta\Omega$ constant and changing the coupling rate $K$ for different coupling phases $\alpha$. The results presented in Fig. \ref{fig_beat_alpha} show that the effect of the coupling rate on the laser frequency difference depends strongly on the coupling phase. For purely dissipative coupling, we observe a monotonically decreasing $\Delta\omega$ as the coupling rate increases in accordance to the Sakaguchi-Kuramoto model,
 \begin{equation}\label{eqn_sqrt}
     \Delta\omega = \Delta\Omega\sqrt{1-(\frac{2K\cos\alpha}{\Delta\Omega}})^2,
 \end{equation}
until phase locking is achieved at the critical coupling. Introducing an increasing coupling phase gradually increases the frequency difference (greater repulsion), eventually causing $\Delta\omega$ to monotonically increase, before collapsing to zero when the critical coupling is reached. For the case of $\alpha = \frac{\pi}{2}$, no stable phase locked solution exists for Eq. (\ref{eqn_freq_lock}), and we observed no locking even for vanishing detuning, as shown analytically in Sec. IV of \cite{Supplemental}. It was previously demonstrated that purely dispersive coupling can phase lock lasers if their amplitudes are not equal\cite{arwas2022anyonic}.  

This frequency repulsion phenomenon is absent in the Sakaguchi-Kuramoto model, and is induced by amplitude dynamics resulting from the dispersive part of the coupling. We study this phenomena perturbatively and numerically in Sec. VII of \cite{Supplemental}, and show that relative amplitude modulations which are correlated to the instantaneous phase difference arise in the system and result in an average frequency repulsion. Numerical simulations of the experimental system with matching parameters are shown in dashed curves in Fig. \ref{fig_beat_alpha}, in good agreement with the experimental results. Similar frequency repulsion effects were previously observed in ring laser gyros \cite{aronowitz1977positive,faucheux1988ring}. 

\begin{figure}[h]
    \centering
    \includegraphics[width=1\linewidth]{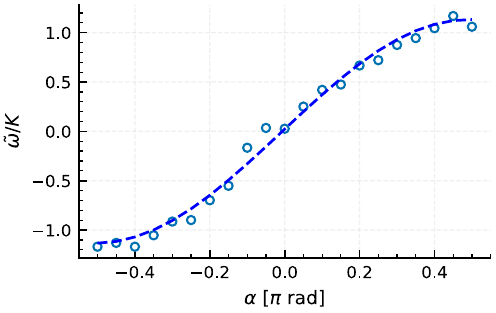}
    \caption{The common frequency of two locked lasers as a function of coupling phase $\alpha$, for $K=1$MHz, $\Delta\Omega=0$. The coupling phase causes a shift of the common frequency of the locked system. The dashed line is a fit to $y=a\sin(x+b)$, as suggested by Eq. (\ref{eqn_freq_shift}), showing a good fit to theory with $a=1.13, b=0.02$.}
    \label{fig_lock_alpha}
\end{figure}

Next we investigate the effects of complex coupling on the common frequency of phase locked lasers. When the lasers are phase locked, they oscillate at a common frequency $\tilde{\omega}\equiv \dot{\phi}_1 = \dot{\phi}_2$, given by
\begin{equation}\label{eqn_freq_shift}
\begin{split}
    &\tilde{\omega} =  \tilde{\Omega} + \frac{1}{2}K\cos\alpha(\frac{A_2}{A_1} - \frac{A_1}{A_2})\sin(\phi_1-\phi_2)\\
    &+\frac{1}{2}K\sin\alpha(\frac{A_2}{A_1} + \frac{A_1}{A_2})\cos(\phi_1-\phi_2),\\
    \end{split}
\end{equation}
where $\tilde{\Omega}\equiv \frac{\Omega_1+\Omega_2}{2}$ is the average natural frequency of the two lasers. In the case of dissipative coupling, since $\alpha=0$ and since the lasers tend to equalize amplitudes, both the second and third terms in Eq. (\ref{eqn_freq_shift}) will vanish and the lasers will oscillate at their average resonant frequency. When $\alpha \ne 0$, neither term will generally vanish, resulting in a frequency shift proportional to $K$. This shift can result in the lasers oscillating in a frequency that is higher (or lower) than either of their resonant frequencies. Figure \ref{fig_lock_alpha} shows the measured oscillating frequency of mutually coupled lasers with $\tilde{\Omega}=\Delta\Omega=0$ and $K=1$MHz for varying coupling phase, by measuring their beating frequency with a known reference \cite{Supplemental}. The results show a good fit to Eq. (\ref{eqn_freq_shift}), under assumption of $A_1=A_2$ (which is suggested from symmetry).

\section{Enhanced synchronization with complex coupling}
\begin{figure}[h]
    \begin{subfigure}{1\linewidth}
    \centering
    % \hspace{5cm}
    \adjustbox{trim=0cm 0 0.95cm 0, clip}{ % Left Bottom Right Top
        \includegraphics[width=1\linewidth]{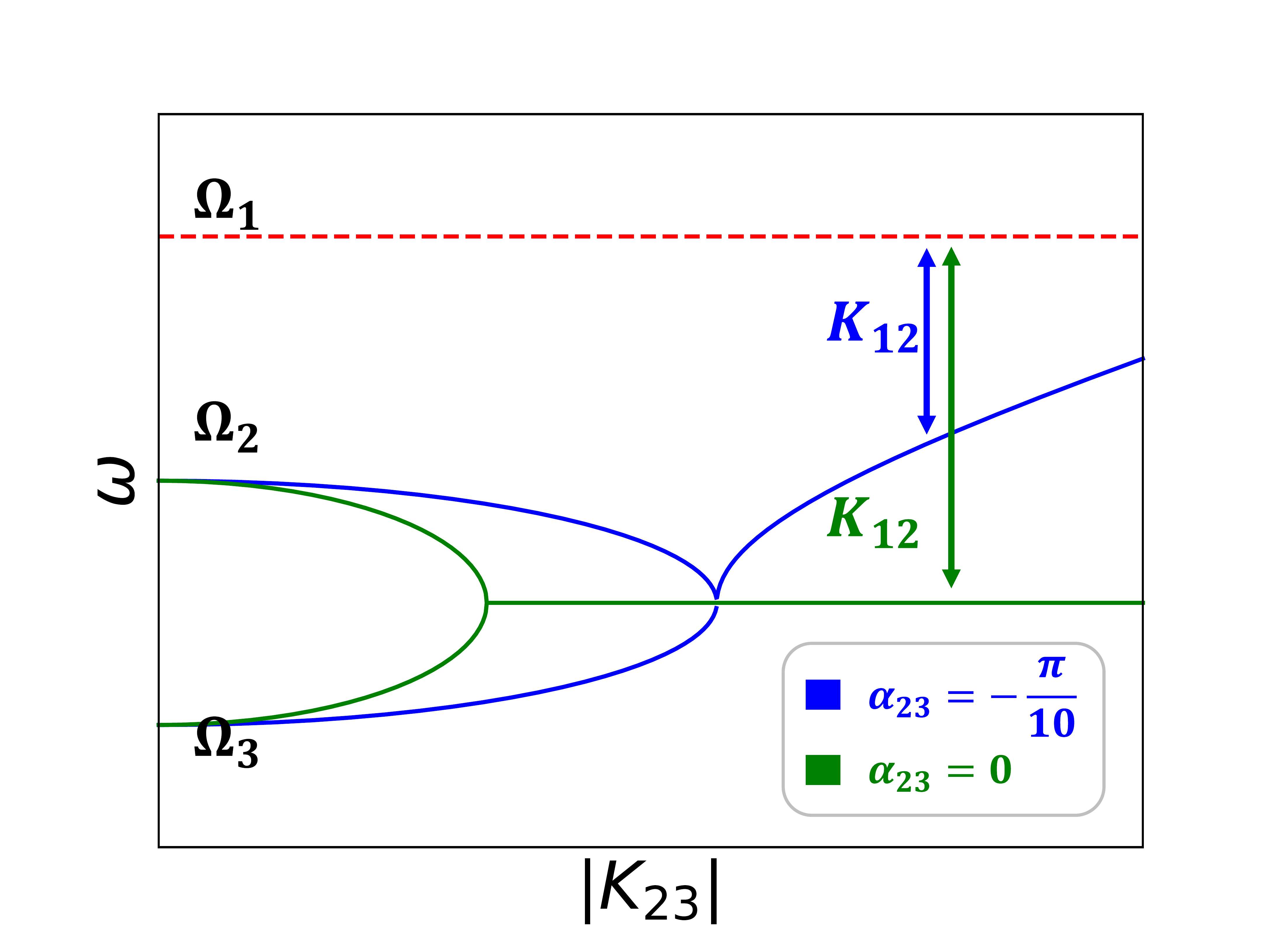}
    }
    \end{subfigure}

    \begin{subfigure}{\linewidth}
    \centering
    \includegraphics[width=1\linewidth]{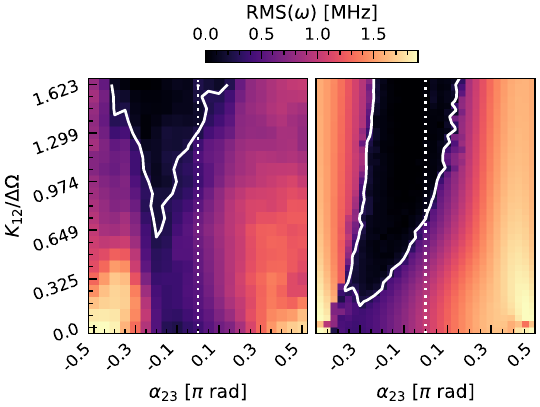}

    \centering
    \end{subfigure}
    
    \caption{Enhanced coherence using complex coupling. \textbf{Top:} Schematic illustration of the experimental scheme, showing the frequency of each laser as a function of the coupling rate $\abs{K_{23}}$ and phase $\alpha_{23}$. The frequency shift induced by complex coupling (blue curve) reduces the frequency difference between the third laser and the rest of the system, when compared to the case of dissipative coupling (green curve). \textbf{Bottom:} Experimental (left) and simulated (right) coherence of the three laser system measured by the RMS of the beating spectrum. The system is completely synchronized when RMS $ = 0$, highlighted by the solid white line. The white dashed marks $\alpha_{23}=0$. The required coupling for synchronization is minimal for $\alpha_{23}\approx -0.28\pi$.}
    \label{fig_enhenced_coup}

\end{figure}
We now consider whether the locked frequency shift that is induced by complex coupling can be exploited for enhancing synchronization of multiple lasers. We employ an array of three lasers and use the following scheme: We couple a pair of lasers with complex coupling, so as to shift their common frequency towards the third laser, thereby decreasing the frequency difference between them and reducing the total coupling required to synchronize all three, as illustrated in the top panel of Fig. \ref{fig_enhenced_coup}.

We performed an experiment with an array of three lasers having frequency detunings of $\Omega_{1,2,3}=-\frac{1}{2}\Delta\Omega,0,\frac{1}{2}\Delta\Omega$, respectively, where $\Delta\Omega = 1.5$ MHz. In general, there are three coupling terms in the array (one between each pair of lasers), each defined by two parameters (coupling rate $K_{ij}$ and phase $\alpha_{ij}$). We expect that for nonzero $\alpha_{23}$, the lasers can be synchronized with a reduced value of $K_{12}$ and hence reduced total coupling. To confirm, we set $K_{13}=0$, $\abs{K_{23}} = 1.22$ MHz, and $\alpha_{12}=0$, such that we only vary two parameters: $\abs{K_{12}}$ and $\alpha_{23}$. The results of the experiment are presented in the bottom left panel of Fig. \ref{fig_enhenced_coup}, where for each choice of parameters we measure the degree of phase locking in the system, quantified by the total RMS of the beating spectrum of the three lasers. Low RMS indicates a highly coherent system where lasers share the same frequency. It is evident that for $\alpha_{23} < 0$ coherence is improved and the coupling rate needed for complete synchronization is reduced from $\frac{K_{12}}{\Delta\Omega} \simeq 1.6$ for purely dissipative coupling, to $\frac{K_{12}}{\Delta\Omega}\simeq 0.7$ for $\alpha_{23}\approx -0.28\pi$. The corresponding numerical simulations shown in the bottom right panel are in agreement with experimental result, although they differ quantitatively due to experimental noise.

\section{Conclusions}
We investigated the effects of complex coupling on synchronization (phase locking) of coupled lasers. Our results show that complex coupling induces a mix of dissipative and dispersive dynamics, promoting both synchronicity through frequency attraction or decoherence through frequency repulsion, depending on the system parameters. We have also shown that while complex coupling requires larger coupling rates to synchronize two lasers compared to purely dissipative coupling, its effects can be exploited to enhance synchronization in a tailored system of three lasers, where we have shown a factor of 2 reduction in the amount of coupling required for synchronization. Our results are supported by numerical simulations, and diverge from the predictions of phase oscillator models, indicating that the amplitude degree of freedom of the lasers plays an important role in the observed behavior.

The high degree of reconfigurability of our digital cavity arrangement enables precise control over geometry, detuning, and coupling phases. It is an ideal testbed for exploring complex phenomena that are difficult with fixed-architecture fabricated devices. In particular, the dynamics of many of those platforms are well simulated by complex coupling, such as the effects of the linewidth enhancement factor in semiconductor lasers\cite{nair2021using} or dispersive-dissipative interactions in polariton condensates\cite{moroney2021synchronization}. Looking forward, our platform could be used to explore the effects of complex coupling in engineered laser networks\cite{vigne2026tailoring}, investigate the role of disorder in non-Hermitian lattices and optimize coherence via engineered dispersion in large laser arrays\cite{manannikov2026klein}.

\section{Acknowledgments - } The authors acknowledge Nathan Vigne and Hui Cao for helpful discussions. The authors also acknowledge the support from the Minerva Stiftung, with funding from the Federal German Ministry for Education and Research.

\section*{Appendix}
\subsection{Coupling via FF SLM}
We consider some field distribution in the cavity $E(x,y)$ at the NF plane of the cavity, and apply some mask $M(k_x,k_x)$ on the FF SLM, where $k_{x,y} = \frac{x,y}{\lambda f}$ are the coordinates in the Fourier plane of the cavity and $f$ is the focal length of the lens $L_1$ in Fig. \ref{fig:exp_setup}. In the Fourier plane, the field distribution is given by
\begin{equation}
    \tilde{E}'(k_x,k_y) =\tilde{E}\cdot M,
\end{equation}
where $\tilde{E}= FT[E]$, the Fourier transform of the original field distribution. In the subsequent NF plane of the cavity,
\begin{equation}
    E'(x,y) = FT[\tilde{E}'] \propto E(-x,-y) \ast \tilde{M},
\end{equation}
where we used the properties of the Fourier transform and the convolution theorem. Hence, the coupling function is determined by the Fourier transform of the FF SLM mask.

For coupling two lasers separated by distance $d$ in the $x$ direction, we apply a mask of the form
\begin{equation}
    M(k_x,k_y) = \frac{1}{1+2K} + \frac{2K}{1+2K}e^{i\alpha}\cos(\frac{x d}{\lambda f}),
\end{equation}
where the $K$-dependent factors are chosen to provide a coupling rate $K$ with phase $\alpha$, and $\abs{M}\le 1$.
The mask yields a convolution kernel
\begin{equation}\label{eqn_coupling_kernel}
    \tilde{M}(x,y) = \frac{1}{1+2K}\delta(x,y) + \frac{K}{1+2K}e^{i\alpha}(\delta(x-d) + \delta(x+d))
\end{equation}

The coupling between the lasers is defined as the relative injection between them, which can approximated by the ratio of terms in Eq. (\ref{eqn_coupling_kernel}) as
\begin{equation}\label{eqn_coupling_simp}
    K(x,y) \approx \Bigg{\{}\mqty{Ke^{i\alpha} && x=\pm d \\ 0 && \text{otherwise}.}
\end{equation}

A more precise evaluation of the coupling done via numerically calculating the overlap between $E$ and $E'$ agrees with Eq. (\ref{eqn_coupling_simp}) to within a few percent \cite{fabiny1993coherence}. Since the Fourier transform is a linear operation, different masks are additive, allowing to easily generate more elaborate coupling schemes (such as those used in Sec. IV of the main text) by summing their required masks and re-normalizing the coefficients accordingly.

\bibliography{main}

\FloatBarrier

\clearpage
\newpage
\widetext
\begin{center}
\textbf{\Large Supplemental Materials}
\end{center}
\setcounter{figure}{0}
\renewcommand{\figurename}{Fig.}
\renewcommand{\thefigure}{S\arabic{figure}}
\setcounter{section}{0}
\section{Detailed Experimental Arrangement}
\begin{figure}[h]
    \centering
    \includegraphics[width=1\linewidth]{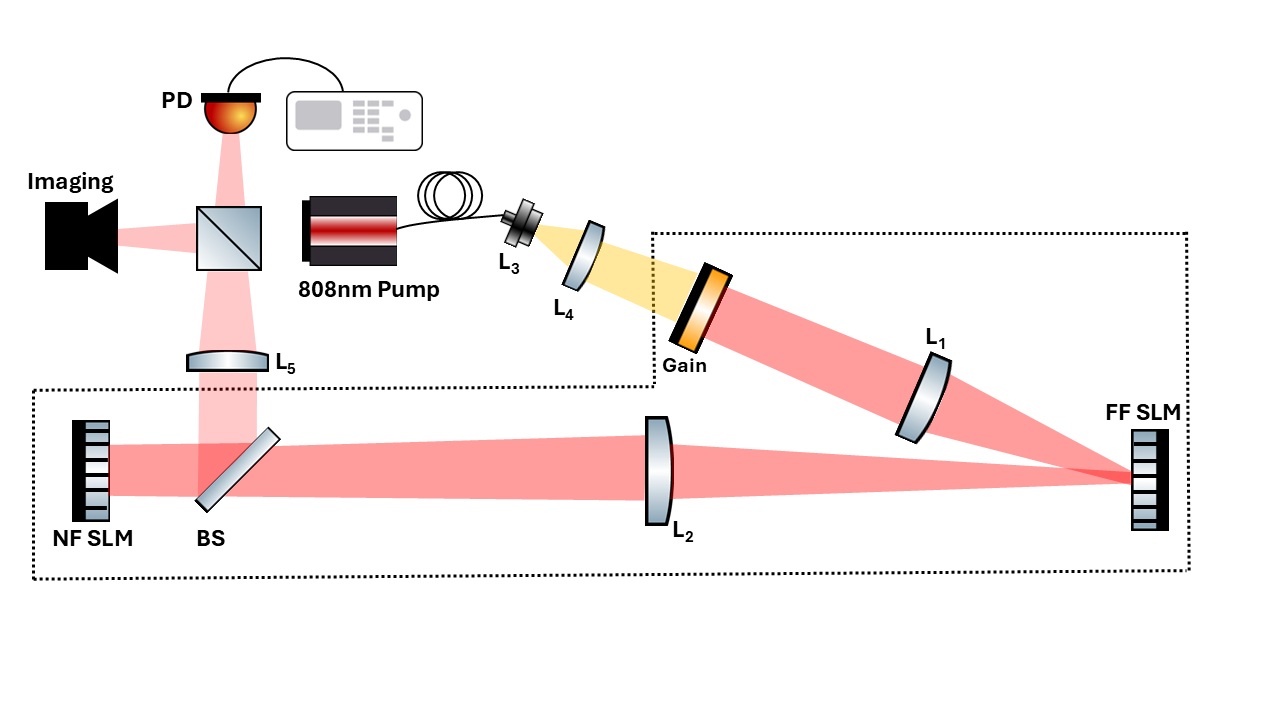}
    \caption{Experimental arrangement of a digital degenerate cavity laser. The cavity contains a Nd:YVO4 gain medium with an HR coated backplane, two achromatic doublet lenses labeled $L_1,L_2$, two spatial light modulators (NF SLM and FF SLM) and a beam sampler (BS).The gain medium is pumped with a $808$nm diode laser coupled to a multimode fiber whose output is imaged onto the gain by lenses $L_3,L_4$. The cavity output is imaged with lens $L_5$ onto an imaging camera and an amplified photodiode, labeled $PD$.}
    \label{fig_sup_exp_setup}
\end{figure}

Our experimental arrangement, shown in Fig. \ref{fig_sup_exp_setup}, consists of a degenerate cavity laser: The cavity hosts a folded $4f$ telescope and a Nd:YVO4 gain medium ($3m\text{m}$ width and $5m\text{m}$ diameter, fluorescence time $\tau_f\approx 90\mu\text{s}$). The cavity is defined by two reflective surfaces acting as end mirrors: A reflective spatial light modulator (Hamamatsu LCOS-SLM X151213, labeled NF SLM), and the back surface of the gain medium coated in high reflectivity coating ($T<1\%$). Two intracavity lenses (achromatic doublets, $L_1,L_2$ with $f_1=20$cm,$f_2=75$cm) in a $4f$ configuration form an intracavity telescope. An additional SLM (Holoeye PLUTO-2.1 NIR-149, labeled FF SLM) is placed in the shared focal plane of the two lenses, acting as a folding mirror and a digitally controlled phase and amplitude mask.

The pump laser is a $808n\text{m}$ multimode diode laser coupled to a multimode fiber with a maximum pump power of $150W$. The fiber output is imaged onto the gain medium such that the pump intensity profile is a top-hat with a $2$mm diameter. The pump laser is operated in quasi-CW operation, with pulse duration of $400\mu \text{s}$ and a repetition rate of 4Hz, with pump power which is roughly double the threshold pump power $P_{th}$. The cavity emission is polarized to be compatible with the SLM polarization.

The cavity is sampled using a thin-film beam sampler with $R\sim 4\%$ and imaged onto a camera and an amplified photodiode with a $380$MHz bandwidth.
We note that since the bandwidth of the gain medium ($\approx 50$GHz) is much larger than $\nu_{FSR}$ our cavity supports many longitudinal modes, that serve as independent experimental realizations which are averaged over in our measurements\cite{chriki2018spatiotemporal,pal2020rapid}.

The formation of individual laser spots and the coupling between them is controlled using the intra-cavity SLMs, where the phase of adjacent blocks of $2\times 2$ pixels are manipulated to control the local effective phase and amplitude of the displayed mask pattern \cite{tradonsky2021high}. By applying this technique to the NF SLM, we create gaussian laser spots and control their relative detuning\cite{pando2023improved}. We then apply the same technique to the FF SLM to control the coupling rate and phase between the lasers as detailed in Appendix A.

\section{Beating frequency measurement method}
The individual laser frequencies are determined by resorting to heterodyne measurement. Specifically, the cavity output's far field is imaged onto an amplified photodiode with a 380 MHz bandwidth (Thorlabs PDA015-C2). The magnification ratio (determined by lens $L_5$ in Fig. \ref{fig_sup_exp_setup}) is set in accordance to the laser distance, such that the interference fringe size is roughly twice the diameter of the photodiode active area, to ensure optimal detection of the beating signal. 

This heterodyne method can only measure the frequency difference between the lasers and not their absolute frequencies. For measurements of the absolute frequency shift (such as for Fig. 4) we introduce an additional laser as a stable reference against which the other signals are compared. This reference laser shares the same degenerate cavity as the other lasers, thus minimizing relative frequency fluctuations due to thermal effects or mechanical vibrations. To decouple the reference from the lasers, it is displaced vertically from the measured lasers which are separated horizontally from each other, while coupling is applied only on the horizontal axis. 

As stated before, our laser supports many longitudinal modes, all interfering on the photodiode. This creates many copies of the beating signal across the spectrum separated by $\nu_{FSR}$, which we average together to increase the signal to noise ratio. 

\section{Stability of equal amplitude fixed point with dissipative coupling}
In this section we show that the two mutually coupled lasers with dissipative coupling, the equal amplitude state is a stable fixed point of the system.

We consider two coupled lasers described by the laser rate equations (Eq. (1a-1c)). We assume purely dissipative coupling ($\alpha=0$), equal pumping and losses for both lasers ($P_1=P_2=P; \gamma_1=\gamma_2=\gamma)$, and denote their relative detuning by $\Delta\Omega = \Omega_1-\Omega_2$. We also assume without loss of generality that $K>0$.

For the case of purely dissipative coupling, all equations are symmetric under the substitution of $A_1\leftrightarrow A_2$. Therefore, we assume the system settles into a phase locked state with
$$ A_1 = A_2 = A_0; \quad G_1 = G_2 = G_0; \quad \Delta\phi = \phi_2-\phi_1 \equiv \Delta\phi_0, $$
where we used the fact that at steady state, $G_i$ is a function of only $A_i,P_i$. Note that phase locking requires $-\pi/2 \le \Delta\phi_0 \le \pi/2$.

To evaluate the linear stability of the state, we introduce a small perturbation,
$$\delta A \equiv A_1-A_2; \quad \delta G \equiv G_1-G_2; \quad \delta\phi \equiv \Delta\phi - \Delta\phi_0.$$

Plugging the perturbation into Eq. (1a)-(1c) and keep terms to first order, we derive the Jacobian matrix,
\begin{equation}
    \dv{t}\mqty(\delta A \\ \delta G \\ \delta \phi) =
    \mqty(\frac{1}{\tau_c}(G_0-\gamma) - K\cos\Delta\phi_0 & \frac{1}{\tau_c}A_0 & 0 \\
    -\frac{2}{\tau_f}G_0A_0 & -\frac{1}{\tau_f}(1+A_0^2) & 0 \\
    0 & 0 & -2K\cos(\Delta\phi_0))
    \mqty(\delta A \\ \delta G \\ \delta \phi)
\end{equation}

The Jacobian is block diagonal, and the phase perturbation is decoupled from the amplitude and gain perturbations. Since $\cos\Delta\phi_0>0$ for all allowed values of $\Delta\phi_0$, the phase perturbation decays exponentially back to its stable value.

We are left with a $2\times 2$ Jacobian matrix for the amplitude and gain perturbations. The fixed point is stable if both eigenvalues have negative real parts. For that purpose, it is enough to show that the Jacobian has a negative trace and positive determinant.

We use the fact that at the fixed point, we can use Eq. (1a) to get
\begin{equation}
    \frac{1}{\tau_c}(G_0-\gamma) = -K\cos\Delta\phi_0,
\end{equation}
such that the trace of the Jacobian is given by
\begin{equation}
    Tr = -2K\cos\phi_0 - \frac{1}{\tau_f}(1+A_0^2).
\end{equation}
Each term is negative so their sum is strictly negative.

The determinant is given by
\begin{equation}
    \Delta = 2K\cos\Delta\phi_0 \cdot \frac{1}{\tau_f}(1+A_0^2) + \frac{2}{\tau_f}G_0A_0^2,
\end{equation}
which is strictly positive.

We conclude that the equal amplitude phase locked state is linearly stable. We note that if the detuning was larger than the critical coupling such that the lasers were not phase locked, the $\cos\Delta\phi_0$ would oscillate over time but the argument would still hold on average.

Lastly, we also show that in the case of complex coupling, the equal amplitude state is generally not valid: It is sufficient to consider Eq. (1a) when $\alpha\ne 0 $ assuming an equal amplitude phase locked state,
\begin{equation}
    0 = \frac{1}{\tau_c}(G_0-\gamma)A_0 + KA_0\cos(\pm\Delta\phi_0 +\alpha),
\end{equation}
where the sign of $\Delta\phi$ in the cosine term changes between the different lasers. Subtracting the equations for the two lasers yield
\begin{equation}
    0 = 2KA_0\sin\sin\Delta\phi_0,
\end{equation}
which implies $\Delta\phi_0 = 0$.

However, Eq. (2) requires
\begin{equation}
    \dot{\Delta\phi} = 0 = \Delta\Omega - 2K\cos\alpha\sin\Delta\phi_0 = \Delta\Omega,
\end{equation}
which means that for complex coupling, the equal amplitude state is only possible in the vanishing detuning case.

\section{Stability analysis of phase locking with purely dispersive coupling}
In this section we show that in the case of purely dispersively coupled lasers with equal pumping rates and losses, the phase locked state is never a stable fixed point.

Similarly to the last section, we use the same identical pump and loss assumption as the previous section, but in this case take the coupling as purely dispersive ($\alpha = \pi/2$).

The phase difference equation (Eq. (2)) reads
\begin{equation}
    \dot{\Delta\phi} = \Delta\Omega - K(\frac{A_1}{A_2} - \frac{A_2}{A_1})\cos\Delta\phi
\end{equation}

We assume there exists a phase locked state such that
$$\dot{\Delta\phi} = 0,$$
and introduce a perturbation to the phase difference,
$$\Delta\phi = \Delta\phi_0+\delta\phi.$$

Expanding around the perturbation and keeping terms to first order we get
\begin{equation}\label{eqn_disp_stability}
    \dot{\delta\phi} = [K(\frac{A_1}{A_2}-\frac{A_2}{A_1})\sin\Delta\phi_0]\delta\phi.
\end{equation}

This perturbation is unstable if the term in the square brackets is positive. To show this, we use Eq. (1a) for the unperturbed phase locked state, and subtract the equations for the two lasers
\begin{equation}\label{eqn_gain_difference}
    G_1-G_2 = -\tau_c K (\frac{A_1}{A_2} + \frac{A_2}{A_1})\sin\Delta\phi_0
\end{equation}

From the gain dynamics described by Eq. (1c) and using the equal pump assumption, we can determine that $G_i$ decreases monotonically with $A_i^2$.  Therefore, the gain difference $G_1-G_2$ has the opposite sign of $A_1-A_2$ and $\frac{A_1}{A_2} - \frac{A_2}{A_1}$.

We assume without loss of generality that $A_1>A_2$, and hence $G_1<G_2$. As a result, the LHS of Eq. (\ref{eqn_gain_difference}) is negative, meaning that $\sin\Delta\phi_0$ must have the opposite sign (and the same sign as $A_1-A_2$). Plugging this into Eq. (\ref{eqn_disp_stability}) we see that the term in the square brackets is indeed positive, implying that the perturbation grows exponentially and that the phase locked state is unstable, as required.

In the above proof we only considered a perturbation to the phase difference. We note that perturbing the amplitude and gain variables in addition to the phase difference does not affect the conclusion: It can be shown that even in the case where all variables are perturbed, the Jacobian matrix is a triangular block matrix. As a result, the coefficient of the phase equation (which we found above to be positive) is an eigenvalue of the full Jacobian matrix, implying that it is at least a saddle point and not a stable fixed point.

\section{LRE simulation details}
The numerical simulations of the full LRE  (Eqs. (1a-1c)) were performed with a standard numerical ODE solver using a Runge-Kutta method. The parameters were chosen to mimic experimental parameters in normalized units and presented in table \ref{tab:sim_parameters} below. The simulations are initialized in cold cavity conditions, with $G=0$ and small random fields. Quantities of interest are calculated and averaged over $\tau_f/\tau_c$ time steps. Each simulation was performed with $10$ randomly chosen initial conditions, and the final results were averaged over all realizations. The coupling rate and phase are reciprocal in all simulations $K_{ij} = K_{ji}, \alpha_{ij}=\alpha_{ji}$ and are written explicitly for all results.

\begin{table*}[h]
\centering
\caption{Numerical parameters used in LRE simulation throughout the paper}\label{tab:sim_parameters}

\renewcommand{\arraystretch}{1.18}
\setlength{\tabcolsep}{6pt}
\begin{tabular}{p{0.4\textwidth}p{0.4\textwidth}}
\hline\hline
Parameter & Value \\
\hline
Cavity roundtrip time & $\tau_c=1$ \\
Gain Fluorescence lifetime & $\tau_f = 6000$ \\
Laser loss & $\gamma = 0.25$ \\
Gain saturation intensity & $I_{sat} = 1$ \\
Pump parameter & $P = 2$ \\
Initial field scale & $E_0 = 10^{-3}$ \\
Initial gain scale & $G_0 = 0$ \\
Total simulation time & $T = 5\tau_f$ \\
Number of random realizations & $N = 10$ \\
Detuning (Fig. 3) & $\{\Omega_1,\Omega_2\} = 0.1,-0.1$ \\
Detuning (Fig. 5) & $\{\Omega_1,\Omega_2,\Omega_3\} = -0.1,0,0.1$\\
\hline\hline
\end{tabular}
\end{table*}

% $\tau_c=1, \tau_f=10^{4},\gamma=0.25,I_{sat}=1$. The detuning and coupling values were chosen in accordance to the specific simulation. The simulation were initialized in cold cavity conditions, i.e. $G=0$ for all lasers, and inital fields $E_i$ were chosen randomly with $\abs{E_i}\sim 10^{-3}$. The simulation time was $T = 5\tau_f/\tau_c$ time steps which is sufficient for the simulation to converge to a steady state. We calculated quantities of interest and averaged over the last $\tau_f/\tau_c$ time steps. Each simulation was averaged over several runs executed with random initial conditions.

\section{Absence of frequency repulsion in Sakaguchi-Kuramoto dynamics}
As noted in the main text, coupled laser dynamics reduce to Sakaguchi-Kuramoto dynamics under the assumption of equal amplitudes $A_n=A$. Under this assumption, Eq. (1b) reduces to
\begin{equation}\label{eqn_sakaguchi}
    \dot{\phi}_n = \Omega_n + K\sin(\alpha)\cos(\Delta\phi_{nm})-K\cos(\alpha)\sin(\Delta\phi_{nm}),
\end{equation}

which is exactly the Sakaguchi-Kuramoto model equation \cite{sakaguchi1988mutual}. We claim that the observed frequency repulsion shown in Fig. 3 in the main text cannot be accounted for by Eq. (\ref{eqn_sakaguchi}). To show this, we consider the frequency difference

\begin{equation}\label{eqn_sakaguchi_delta}
\Delta\omega \equiv \Delta\dot{\phi}_{12} = \Delta\Omega_{12} - 2K\cos(\alpha)\sin(\Delta\phi_{12})
\end{equation}

Equation (\ref{eqn_sakaguchi_delta}) can be solved exactly for $\Delta\omega$ \cite{sakaguchi1988mutual}, yielding 
\begin{equation}\label{eqn_delta_omega}
    \Delta\omega = \sqrt{\Delta\Omega^2 - 4K^2\cos[2](\alpha)}
\end{equation}

Equation (\ref{eqn_delta_omega}) is strictly smaller or equal to $\Delta\Omega$, indicating the observed frequency repulsion of Fig. 3 indeed necessitates the amplitude degree of freedom.

\section{Amplitude dynamics and frequency repulsion: Numerical and analytical evidence}
In this section we provide theoretical and numerical evidence showing that complex coupling induces relative amplitude modulation that results in frequency repulsion.

We resort to a perturbative approach far from the locking regime, $K\ll\Delta\Omega$. Our goal is to write the beating frequency of the lasers given by Eq. (2) and show that an amplitude dependent frequency repulsion term arises.

We define the relative amplitude variable,
\begin{equation}
    x(t) \equiv \ln{\frac{A_1}{A_2}} \rightarrow \frac{A_1}{A_2} = e^x
\end{equation}

and write Eq. (2) as
\begin{equation}\label{eq_2_pert}
    \Delta\dot{\phi} = \Delta\Omega -2K\cos\alpha \cosh{x} \sin\Delta\phi - 2K\sin\alpha\sinh x \cos\Delta\phi
\end{equation}

For class B lasers as those considered in this context, the gain dynamics are much slower than the beating frequency ($\tau_f^{-1} \ll \Delta\Omega$), and their effects are negligible to the first order: The laser amplitudes oscillate faster than the gain can react.

We solve $\Delta\phi$ and $x$ perturbatively to the first order in $K/\Delta\Omega$, and find:
\begin{equation}\label{eq_dphi_pert}
\Delta\phi \approx \Delta\Omega t +\frac{2K}{\Delta\Omega}\cos\alpha \cos(\Delta\Omega t) + \mathcal{O}(K^2/\Delta\Omega^2)
\end{equation}
\begin{equation}\label{eq_x_pert}
 x\approx -\frac{2K}{\Delta\Omega}\sin\alpha\cos(\Delta\Omega t) + \mathcal{O}(K^2/\Delta\Omega^2)
\end{equation}

We see that to first order, dispersive coupling drives relative amplitude oscillations which are out of phase with the phase oscillations, consistent with the numerical simulation results. We plug Eq. (\ref{eq_dphi_pert}-\ref{eq_x_pert}) into Eq. (\ref{eq_2_pert}) and keep terms to first order in $K/\Delta\Omega$ to get
\begin{equation}
    \Delta\dot{\phi}\approx \Delta\Omega - 2K\cos\alpha \big[\sin(\Delta\Omega t) + \frac{2K}{\Delta\Omega}\cos\alpha\cos[2](\Delta\Omega t)\big] + \frac{4K^2}{\Delta\Omega}\sin^2\alpha\cos[2](\Delta\Omega t),
\end{equation}
where we note that the terms in the square brackets are always attractive, pulling the frequencies together, while the last term is a repulsive term, pushing the frequencies away from each other. Taking the time average over a cycle yields
\begin{equation}\label{eq_pert_repulsion}
    \ev{\Delta\dot{\phi}} \approx \Delta\Omega - \frac{2K^2}{\Delta\Omega}\cos(2\alpha).
\end{equation}

Equation (\ref{eq_pert_repulsion}) shows that for $\alpha > \pi/4$, increasing the coupling leads to frequency repulsion, as is observed in our experimental and numerical results. While this analysis is perturbative and only valid in the very weak coupling limit, it reveals that amplitude fluctuations are expected to emerge in our system and to drive the laser frequencies away from each other for large enough dispersive coupling.

\begin{figure}[h]
    \centering
    \includegraphics[width=0.75\linewidth]{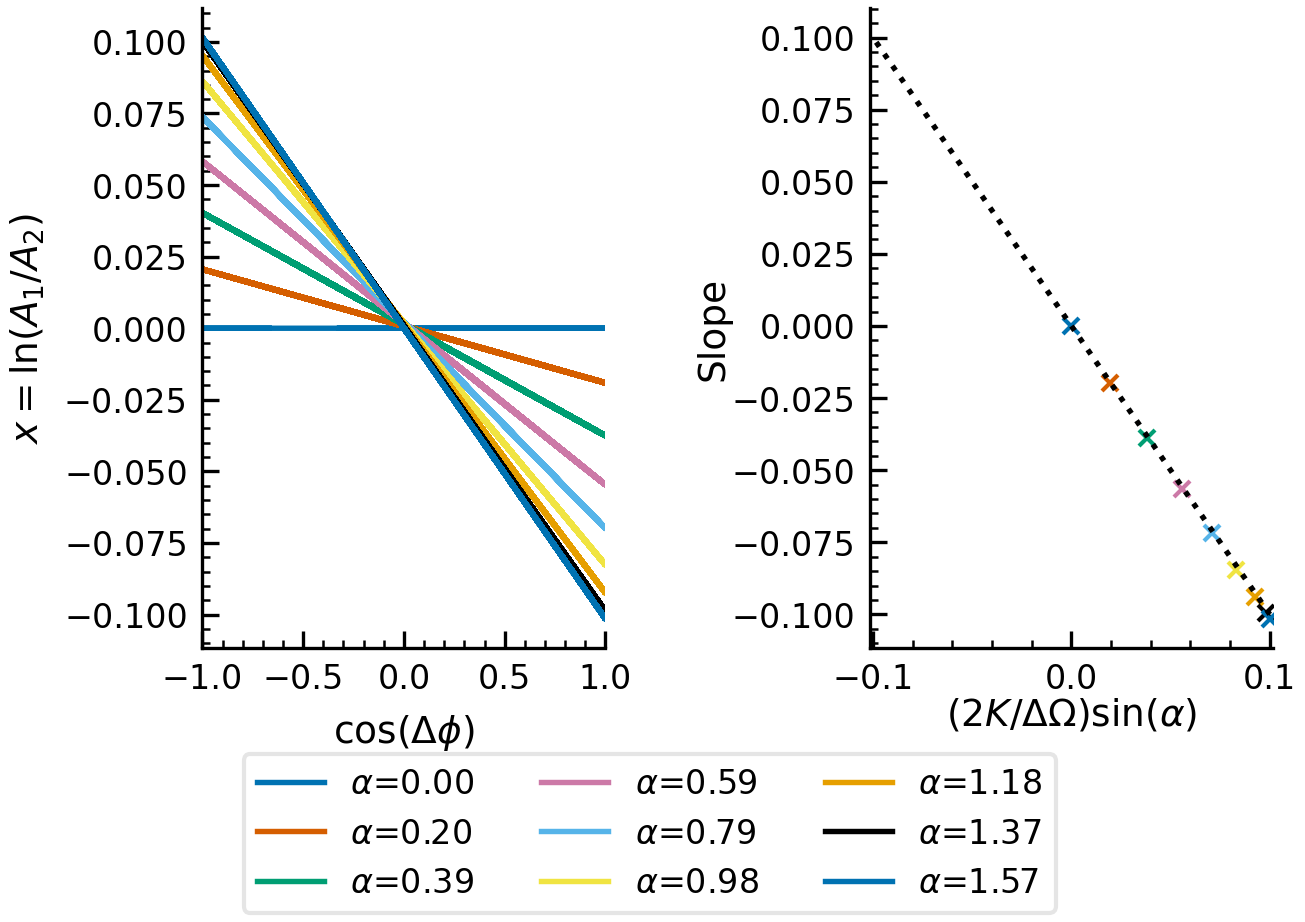}
    \caption{Phase driven relative amplitude modulation in numerically simulated laser rate equations, with $K=0.05\Delta\Omega$. \textbf{Left}: Scatter plot of the relative amplitude of the lasers as a function of the cosine of their relative phase. The modulation depth increases with the coupling phase. \textbf{Right}: The numerically extracted slope of the data in the left plot compared with the theoretically expected value of Eq. (\ref{eq_x_pert}), showing excellent agreement. Dotted line indicated $y=-x$.}
    \label{fig_amp_osc_theory}
\end{figure}

To validate our analysis, we numerically simulated a system of two complex coupled lasers and observed their amplitude and phase dynamics. Figure \ref{fig_amp_osc_theory} shows the steady state results for $K=0.05\Delta\Omega$. The left panel of Fig. \ref{fig_amp_osc_theory} shows the relative amplitude as a function of the cosine of the lasers phase difference, revealing a linear relation in accordance to Eq. (\ref{eq_x_pert}). The right panel of Fig. \ref{fig_amp_osc_theory} compares the slope of the linear trend to the expected slope from Eq. (\ref{eq_x_pert}), showing excellent agreement.

\begin{figure}[h]
    \centering
    \includegraphics[width=0.75\linewidth]{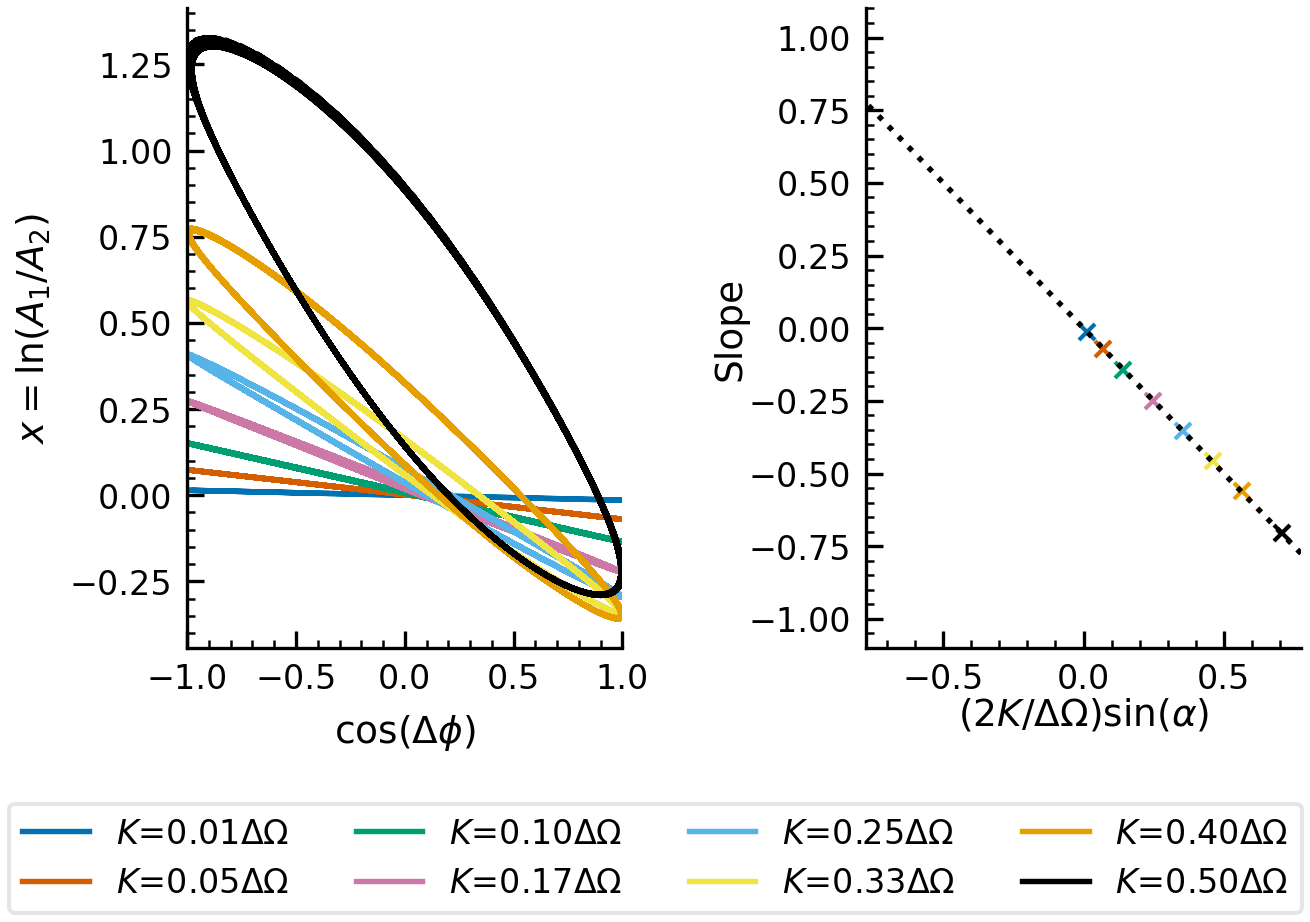}
    \caption{Phase driven relative amplitude modulation in numerically simulated laser rate equations, with $\alpha=\pi/4$. \textbf{Left}: Scatter plot of the relative amplitude of the lasers as a function of the cosine of their relative phase. Increasing the coupling rate introduces additional modulation beyond the perturbative linear result of Eq. (\ref{eq_x_pert}). \textbf{Right}: The numerically extracted average slope of the data is still in agreement with Eq. (\ref{eq_x_pert}). Dotted line indicated $y=-x$.}
    \label{fig_amp_osc_theory2}
\end{figure}

Figure \ref{fig_amp_osc_theory2} shows the numerical results for constant coupling phase ($\alpha=\pi/4$) and increasing coupling. It is evident that as the coupling increases, the linear relation between the relative amplitude modulation and $\cos\Delta\phi$ breaks, and a limit cycle develops. Nevertheless, the average slope is still consistent with the result of Eq. (\ref{eq_x_pert}). It is also interesting to note that the modulation is now strongly asymmetric, presumably due to higher order terms. The direction of asymmetry is reversed upon changing the sign of the coupling phase or the detuning difference.

These results provide insight into the mechanism of frequency pushing:  $x$ and $\cos\Delta\phi$ are negatively (positively) correlated for positive (negative) relative detuning. Therefore, the dispersive coupling term which scales as $\sinh{x}\cos\Delta\phi$ in Eq. (\ref{eq_2_pert}) will have the same sign as the detuning, resulting in frequency repulsion. This is in contrast to the dissipative term, whose $\cosh x$ term is strictly positive. As a result, the amplitude dynamics only effectively modify the magnitude of the dissipative term, and not its sign. 

\end{document}